\documentclass[conference]{IEEEtran}

\def\done{\hspace*{\fill} \rule{1.8mm}{2.5mm} \\ }
\usepackage{cite}

\ifCLASSINFOpdf

\else
\usepackage[dvips]{graphicx}
\graphicspath{{figure/}} \fi

\usepackage[cmex10]{amsmath}

\usepackage{subfigure}
\usepackage{stfloats}
\usepackage{url}

\def\Mu{{\boldmath \mu}}

\def\S{{\textbf S}}
\def\b0{{\bf 0}}

\def\bc{{\bf c}}
\def\bd{{\bf d}}
\def\bx{{\bf x}}
\def\by{{\bf y}}

\def\bz{{\bf z}}

\def\bu{{\bf u}}
\def\bv{{\bf v}}
\def\bw{{\bf w}}

\def\bA{{\bf A}}

\def\bG{{\bf G}}
\def\bR{{\bf R}}
\def\bX{{\bf X}}
\def\bY{{\bf Y}}
\def\bP{{\bf P}}
\def\bQ{{\bf Q}}

\newcommand\bmgl{{\mbox{\boldmath$\lambda$}}}
\def\mcR{{\mathcal R}}

\def\bmgl{{\mbox{\boldmath$\lambda$}}}
\def\Mu{{\mbox{\boldmath$\mu$}}}
\def\Nu{{\mbox{\boldmath$\nu$}}}
\def\mcJ{{\mathcal J}}
\def\mcS{{\mathcal S}}
\def\mcR{{\mathcal R}}
\def\mcN{{\mathcal N}}
\def\mcM{{\mathcal M}}
\newtheorem {lemma} {\hspace{-0.15in} \textbf{Lemma} } 
\newtheorem {theorem}  {\hspace{-0.15in}\textbf{Theorem}}

\def\bcc{\begin{center}}
\def\ecc{\end{center}}

\def\Min{{\rm minimize}}
\def\Max{{\rm maximize}}

\def\ST{{\rm subject\ to}}
\def\dif{\frac{\tt d}{{\tt d}t}}

\begin{document}
%
\title{ A Large Family of Multi-path Dual Congestion Control Algorithms}




\maketitle

\begin{abstract}

The goal of traffic management is efficiently utilizing network
resources via adapting of source sending rates and routes selection.
Traditionally, this problem is formulated into a utilization maximization problem.
The single-path routing scheme fails to react to
instantaneous network congestion. Multi-path routing
schemes thus have been proposed aiming at improving network
efficiency. Unfortunately, the natural optimization problem to
consider is concave but not strictly concave. It thus brings a huge
challenge to design stable multi-path congestion control algorithms.

In this paper, we propose a generalized multi-path utility
maximization model to consider the problem of routes selection and flow control, and derive a family of multi-path dual
congestion control algorithms. We show that the proposed algorithms
are stable in the absence of delays. We also derive decentralized
and scalable sufficient conditions for a particular scheme when
propagation delays exist in networks. Simulations are implemented using both Matlab and NS2,
on which evaluation of the proposed multi-path dual algorithms is exerted. The comparison results,
between the proposed algorithms and the other two existing algorithms,
show that the proposed multi-path dual algorithms with appropriate parameter settings can achieve a stable aggregated throughput while maintaining fairness among the involved users.
\end{abstract}

%
\IEEEpeerreviewmaketitle

\section{Introduction}\label{overview}
The Transmission Control Protocol and the Internet Protocol, known
as TCP/IP, are widespread for guiding traffic flows in the Internet.
In a packet-switch network, a route is computed and selected to send
packets from a source user to a destination user, and the sending
rate is determined by TCP. Traditionally, a single-path routing
scheme is deployed, where the shortest path is chosen by IP routing
in terms of hop count or distance, and the flow rate is varied
according to congestion level along that path.
Ideally, both routes and flow rates should be guided to guarantee
the efficiency and fairness in link bandwidth utilization. There
thus has long been a desire to direct routes selection and rates
variation according to congestion level. However, studies, \emph{e.g.}
\cite{Intro-Earlystudy}, have shown that making paths selection
consistent with congestion level may result in network occlusions
and routing instability. Despite that IP routing is highly scalable,
the static or single-path routing scheme fails to react to
instantaneous network congestion.

Motivated by the applications in ad-hoc networks and overlay TCP,
recently there have been more interests in multi-path routing
scheme \cite{hoc_Crowcroft} \cite{hoc_Johnson}. In that scheme, packets belonging to the same
source-destination pair are transmitted along several routes between
them instead of a single path. Notice that these routes might not be
disjoint. In order to take advantage of multi-path routing scheme,
network users prefer to select the best path among the available
routes in terms of high throughput or low latency. However,
Wang et al. \cite{Intro-Instability} has shown the instability arose from such
interaction between network users and providers, causing barriers to
the deployment of this dynamic routing scheme in packet-based
networks.

Many researchers devote to find a protocol that can be implemented
in a decentralized way by source and routers, and controls the
system to a stable equilibrium point which satisfies some basic
requirements: high utilization of resources, small queues, and a
degree of control over resource allocation. All of these are
required to be scalable, {\em  i.e.}, hold for an arbitrary network,
with possibly high capacity and delay.

A {\em major} difficult for the multi-path congestion control is that the natural
optimization problem to consider is concave but not strictly
concave. It makes that there are possible existence of multiple
equilibriums. Thus, researchers resort to the duality of the primal problem.
One attractive consequence of the dual algorithm is that they
naturally have equilibrium points which make full use of the limited
bandwidth available, while still achieving a notion of fairness
between users. Voice \cite{Voice07TonDual} is the first considering
the stability of multi-path dual algorithm. The method that
extended a single-path result to the multi-path case was used in
\cite{KellyVoice05} for a primal congestion control algorithm. However, the fairness among different users
is not well considered in \cite{Voice07TonDual}.

In this paper, we propose a generalized multi-path utility
maximization model, which is strictly concave and ensure equilibria
satisfying desirable static properties. Then we derive a family of
multi-path dual congestion control algorithms. We show that
the proposed algorithms are stable in the absence of delays, based on which we
derive decentralized and scalable sufficient conditions for a
particular scheme when propagation delays exist in the networks.

The main contributions of our work can be stated as follows:

1)We propose a generalized multi-path utility maximization model, which can reduce to specific models with different parameter settings.
A family of multi-path dual congestion control algorithms derived from the above model can both fully utilize resources under limitation and achieve stability
in the presence of propagation delays in network, while maintaining fairness among different users.

2)We implement both rate-based and window-based simulations respectively using Matlab and NS2,
respectively.  To validate the efficiency of the proposed
algorithms, a comparison is made between the proposed algorithm and
the ones in \cite{KellyVoice05} and \cite{Voice07TonDual} under NS2.
Results show that the proposed multi-path dual algorithms outperform
the later ones in optimal and stable aggregated throughput under an
appropriate value of the average window size.

The remainder of this paper is organized as follows. Related work
is briefly reviewed in Section \ref{relatedwork}. In Section
\ref{utilitymodel}, we present the proposed multi-path utility
maximization model. Both stability in the absence of delays and in
the presence of delays are exhibited in Section \ref{gloabl} and
Section \ref{local}, respectively. Following that is the simulation
results in Section \ref{simu}. This paper is finally concluded in
Section \ref{conclusion}.

\section{Related Work}\label{relatedwork}
In recent years theoreticians have developed a framework that allows
a congestion control algorithm such as Jacobson's TCP to be
interpreted as a distributed mechanism solving a global optimization
problem: for reviews see \cite{Kelly98_SOR} and \cite{Srikant04}.
 The framework is based on fluid-flow models, and
the form of the optimization problem makes explicit the equilibrium
resource allocation policy of the algorithm, which can often be
restated in terms of a fairness criterion. And the dynamics of the
fluid-flow models allow the machinery of control theory to be used
to study stability, and to develop rate control algorithms that
scale to arbitrary capacities. The equilibrium and dynamic
properties for the related congestion control algorithm based on
this framework are summarized in Table \ref{tab:EDP_CCA}.

\begin{table*}[!t]
\centering \caption{Equilibrium and dynamic properties for
congestion control algorithms} \label{tab:EDP_CCA} \vspace{4pt}

\begin{tabular}{|c|c|c|c|c|c|c|c|}
\hline                   & \multicolumn{2}{c|}{Single-path Case}

                                                                                                                                                                                       & \multicolumn{2}{c|}{Multi-path Case}\\ \cline{2-5}
      Algorithms        & Primal algorithms                                                   & Dual algorithms                                                                                               & Primal algorithms                                        & Dual algorithms       \\   \hline
       Global Stability$^\natural$ &  \cite{Kelly98_SOR}$^\ddag$                             & \cite{Kelly98_SOR}$^\ddag$, \cite{Low99}$^\dag$, \cite{Paganini02}$^\dag$, \cite{Paganini05}$^\dag$ & \cite{Kelly98_SOR}$^\S$                        & \cite{Kelly98_SOR}$^\S$, \cite{Lin06Ton}$^\ddag$, \cite{Voice07TonDual}$^\ddag$   \\
\hline Local Stability$^\sharp$    &  \cite{Kelly98_SOR}$^\ddag$, \cite{Vinnicombe02}$^\dag$ & \cite{Kelly98_SOR}$^\ddag$, \cite{Paganini01}$^\dag$, \cite{Kelly03}$^\dag$                         & \cite{KellyVoice05}$^\S$, \cite{Han06Ton}$^\S$ & \cite{Voice07TonDual}$^\ddag$     \\
\hline
\end{tabular}\\
\begin{quote}
    $^\dag, ^\ddag$ and $^\S$ denote the equilibrium point solving the original
    problem, the approximation problem and the relaxed problem respectively\\
    $^\natural$ and $^\sharp$ denote the stability in the absence of propagation delays and the one in the presence of propagation delays respectively.
\end{quote}
\end{table*}

These algorithms can be classified into two major groups, {\em  i.e.},
{\em  primal} algorithms and {\em  dual} algorithms. In general, the
equilibrium point of the algorithm solves the primal (or original)
problem, an approximation problem or the relaxed problem (where the
capacity constrain is replaced by penalties) respectively
\cite{Kelly98_SOR}.

For the single-path case, Vinnicombe \cite{Vinnicombe02} derived
decentralized and scalable stability conditions for a fluid
approximation of a class of Internet-like communications networks
operating a modified form of TCP-like congestion control. Dual
algorithms are classed two groups, {\em  i.e.} delay-based and fair
dual algorithm \cite{Kelly03}. The delay-based dual algorithms
allowed a natural interpretation of the link price as either a real
or virtual queueing delay \cite{Kelly98_SOR}, \cite{Low99},
\cite{Paganini02}, \cite{Paganini05}. It was, however, difficult to
reconcile fairness with stability. Kelly \cite{Kelly03} design a
class of fair dual algorithm, which can achieve weighted
$\alpha$-fairness, and have straightforward delay and stochastic
stability properties .

In multi-path case, there are possible existence of multiple
equilibriums because that the natural optimization problem to consider is concave but not strictly
concave. Meanwhile, when one attempts to use a duality
approach, the dual problem may not be differentiable at every point
\cite{Bert99}. To circumvent these difficulties, Lin et al. \cite{Lin06Ton} used ideas
from proximal point algorithms. Han et al. \cite{Han06Ton} modified
the utility function to ensure a unique equilibrium point and
generalized the algorithm for the case of single-path
\cite{Vinnicombe02} to multi-path. Kelly at al. \cite{KellyVoice05}
improved on the results obtained by Han et al. \cite{Han06Ton}, and
present an algorithm with a sufficient condition for local stability
that is decentralized in the stronger sense that the gain parameter
for each route is restricted by the round-trip time of that route.
However, the majority of above research focuses on extensions of the
primal algorithms proposed by Kelly et al. \cite{Kelly98_SOR}, a
class of single-path primal congestion controls. The primal
algorithms exhibit a trade-off between rate of convergence and
bandwidth utilization at equilibrium since that the desired
equilibrium point only solves the relaxed problem.

\section{A Generalized Multi-path Utility Maximization Model} \label{utilitymodel}
\renewcommand{\theequation}{2.\arabic{equation}}
\setcounter{equation}{0}

First we will give the network model and propose a generalized
multi-path utility maximization model. Then approximation error and
dual problem of the generalized model will be presented.

\subsection{Network Model}
We suppose that the network comprises an interconnection of a set of
{\em sources} $\mcS$, with a set of {\em resources} $\mcJ$. Each
source $s\in \mcS$ identifies a unique source-destination pair.
Associated with each source is a collection of {\em routes}, each
route being a set of resources. If a source $s$ transmits along a
route $r$, then we write $r\in s$. For a route $r$, we let $s(r)$ be
the (unique) source such that $r\in s(r)$. We let $\mcR$ denote the
set of all routes. In the following, we use notations $S, J$ and $R$
to denote the cardinalities of sets $\mcS, \mcJ$ and $\mcR$
respectively.

In our model, a route $r$ has associated with it a flow rate
$x_r(t)\ge 0$, which represents a dynamic fluid approximation to the
rate at which the source $s(r)$ is sending packets along route $r$
at time $t$.

For each route $r$ and resource $j\in r$, let $T_{rj}$ denote the
propagation delay from $s(r)$ to $j$, {\em  i.e.} the length of time
it takes for a packet to travel from source $s(r)$ to source $j$
along route $r$. Let $T_{jr}$ denote the propagation delay from $j$
to $s(r)$, {\em  i.e.} time it takes for congestion control feedback
to reach $s(r)$ from resource $j$ along route $r$. In the protocols
to be considered, a packet must reach its destination before an
acknowledgement containing congestion feedback is returned to its
source. Further, we assume queueing delays are negligible. Thus for
all $j\in r$, $T_{rj}+T_{jr}=T_r$, the round trip time for route
$r$.

The notation $a=(b)_c^+$ denotes that $a=b$ if $c>0$ and
$a=\max(0,b)$ if $c=0$. We abuse notations to use
$\|\bx^s\|_{\frac{1}{q}}$ to denote $(\sum_{r\in
s}x_r^{\frac{1}{q}})^q$ and to use $\|\bmgl^s\|_{1-p}$ to denote
$(\sum_{r\in s}\lambda_r^{1-p})^{\frac{1}{1-p}}$ for any $p>1$ and
$q>1$ such that $\frac{1}{p}+\frac{1}{q}=1$.

\subsection{A Generalized Multi-path Utility Maximization Model}
\renewcommand{\theequation}{2.\arabic{equation}}
\setcounter{equation}{0}

A utility function $U_s(y_s)$ is associated to each source $s\in
\mcS$, which is an increasing, strictly concave and continuously
differentiable function of $y_s$ over the range $y_s>0$. And
$U_s(y_s)\to \infty$ if $y_s\to \infty$. As an example, suppose that
\begin{equation}\label{eq:utilityfunction}
U_s(y_s)=\left\{
  \begin{array}{ll}
    w_s\frac{y_s^{1-\alpha}}{1-\alpha}, & \alpha\ne 1 \\
    w_s\log{y_s}, & \alpha=1
  \end{array}
\right.
\end{equation}
for $w_s>0,\alpha > 0$, so that the resource shares obtained by
different sources are weighted $\alpha$-fair \cite{FairMo00}.
When
$w_s=1, s\in \mcS$, the cases $\alpha\to 0, \alpha=1$ and $\alpha\to
\infty$ correspond respectively to an allocation which achieves
maximum throughput, is {\em proportionally fair} or is
{\em max-min fair} \cite{FairMo00}. TCP fairness, in the case where
each source has just a single route, corresponds to the choice
$\alpha=2$ with $w_s$ the reciprocal of the square of the (single)
round trip time for source $s$ \cite{Srikant04}.
Define the demand function $D_s(\lambda_s)=(U'_s)^{-1}(\lambda_s)$,
a continuous, strictly decreasing function. The demand functions
derived from the class of utility functions defined in
\eqref{eq:utilityfunction} is
\begin{equation}\label{eq:demandfunction}
D_s(\lambda_s)= \left(\frac{w_s}{\lambda_s}\right)^{1/\alpha}.
\end{equation}

For the convenience of making analysis, first we introduce routing
matrix to succinctly express the relationships between routes and
resources. Let $A_{jr}=1$ if $j\in r$, so that resource $j$ lies on
route $r$, and set $A_{jr}=0$ otherwise. This defines a 0-1 matrix
$\bA=(A_{jr}, j\in \mcJ, r\in \mcR)$. The aggregate rate for sources
$s$ is $y_s=\sum_{r\in s}x_r$. Since we wish the total network
utility to be high, it is desirable for a congestion control
algorithm to asymptotically solve the classical Kelly formulation:
\begin{equation} \label{eq:MPNUM}
\begin{array}{ll}
\Max_{\bx\ge 0}& \sum\limits_{s\in\mathcal {S}}U_s(\sum_{r\in s}x_r) \\
{\ST} & \bA\bx\le\bc,
\end{array}
\end{equation}
where $\bc=(c_j, j\in \mcJ)$ with $c_j$ being the capacity of
resource $j$. Note, even if $U_s$ is strictly concave, the whole
objective function is not, due to the linear relationship in
$\sum_{r\in s}x_s$.

A generalized model for the multi-path utility maximization problem
\eqref{eq:MPNUM} is to
\begin{equation}\label{eq:NewPrimal}
\begin{array}{ll}
{\rm maximize}_{\bf x\ge 0}&  \sum\limits_{s\in \mcS}U_s(u_s^{q})\\
{\rm subject\ to \ }&
u_s\le\gamma\sum\limits_{r\in s}x_r^{\frac{1}{q}}+(1-\gamma)y_s^{\frac{1}{q}}\\
                       & \sum\limits_{r\in s}x_r=y_s, \ s\in \mcS\\
                       & \bA\bx\le\bc,
\end{array}
\end{equation}
where $\gamma\in [0,1]$ and $q>1$. Given $q>1$,  \eqref
{eq:NewPrimal} reduces to the one proposed by Voice in
\cite{Voice07TonDual} with $\gamma=1$ and to \eqref{eq:MPNUM} with
$\gamma=0$.  The {\em \textbf{motivation}} for such formulation is that \eqref{eq:NewPrimal} can reduce to the classical Kelly formulation \eqref{eq:MPNUM} and the one in \cite{Voice07TonDual} under different parameter settings. The {\em \textbf{advantages}} of formulation \eqref{eq:NewPrimal} are two folds, we can not only provide a direct insight into the reason for the stability of the dual congestion control with respective to previous work in \cite{Kelly98_SOR} and \cite{Low99}, but also avoid choosing sufficient large parameter $p$ to approximate \eqref{eq:MPNUM} with respective to the work in \cite{Voice07TonDual}, which implies large risk of numerical instability.

To ensure that the objective function is strictly
concave, we make {\bf Assumption H}: For each $s\in \mcS, u_s^{q-1}U'_s(u_s^q)$ is a strictly decreasing function of $u_s$.
This is true for the weighted $\alpha-$fairness utility function
\eqref{eq:utilityfunction} if an appropriate $p$ is chosen, such as
$\alpha p>1$, where $p>1$ and $\frac{1}{p}+\frac{1}{q}=1$.

\subsection{Approximation Error}
To solve the non-strict concave of the
objective function in \eqref{eq:MPNUM} with $\bx$, only the $\gamma$ fraction of the
$\frac{1}{q}$ power of aggregate rate $y_s=\sum_{r\in s}x_r$ is substituted by
$\sum_{r\in s}x_r^\frac{1}{q}$ in problem \eqref{eq:NewPrimal}. We can bound
how far the solution to \eqref{eq:NewPrimal} is from maximizing
aggregate user utility.

\noindent\begin{lemma}[Approximation error] Let $(\bx', \by')$ be any
optimal solution to \eqref{eq:MPNUM}, and $(\bx, \by, \bu)$ be
the optimal solution to \eqref{eq:NewPrimal}. We have
\begin{equation}\label{eq:AppErro1}
\sum_{s\in \mcS} U_s\big(\sum_{r\in s} x'_r\big)\ge \sum_{s\in \mcS}
U_s\big(\sum_{r\in s} x_r\big)
\end{equation}
and
\begin{equation}\label{eq:AppErro2}
\sum_{s\in \mcS} U_s\big(e_{\gamma}\sum_{r\in s} x_r\big)\ge \sum_{s\in \mcS}
U_s\big(\sum_{r\in s} x'_r\big),
\end{equation}
where error factor
$e_\gamma=\big(\gamma|s|^{\frac{1}{p}}+(1-\gamma)\big)^{q}$
and $|s|$ denotes the number of route serving for source $s$.
\end{lemma}

\noindent{\bf Proof}:
It's obvious that $(\bx,\by)$ is
feasible for \eqref{eq:MPNUM}. So the inequality \eqref{eq:AppErro1}
is followed by the optimality of $(\bx', \by')$ to \eqref{eq:MPNUM}.

For $p>1, q>1$ and $\frac{1}{p}+\frac{1}{q}=1$, we have $\sum_{r\in
s}x_r^{\frac{1}{q}}\le |s|^{\frac{1}{p}}\big(\sum_{r\in
s}x_r\big)^{\frac{1}{q}}$ with the famous H\"{o}lder inequality
\cite{Luenberg}. Combining it with the facts that
$u_s=\gamma\sum_{r\in s}x_r^{\frac{1}{q}}+(1-\gamma)(\sum_{r\in
s}x_r)^{\frac{1}{q}}$ and $(\cdot)^{q}$ increasing, we have
\begin{equation}\label{eq:AppErro3}
e_\gamma\sum_{r\in s}x_r\ge u_s^{q}.
\end{equation}
Now let $u'_s=\gamma\sum_{r\in s}
{x'_r}^{\frac{1}{q}}+{(1-\gamma)y'_s}^{\frac{1}{q}}$. It's obvious
that $(\bx', \by', \bu')$ is feasible for \eqref{eq:NewPrimal}. By
the optimality of $(\bx, \by, \bu)$ to \eqref{eq:NewPrimal}, we have
\begin{equation}\label{eq:AppErro4}
U_s(u_s^{q})\ge {{U_s({{u}'_s}}}^{q}).
\end{equation}
Since $(\cdot)^{\frac{1}{q}}$ is a subadditive function and
$(\cdot)^{q}$ is increasing, we have
\begin{equation}\label{eq:AppErro5}
{{u'_s}}^q\ge \sum_{r\in s}x'_r
\end{equation}
for $y'_s=\sum_{r\in s} x'_r$. Combing the inequalities
\eqref{eq:AppErro3}, \eqref{eq:AppErro4}, \eqref{eq:AppErro5} and
the fact that $U_s(\cdot)$ is increasing, we get the desired result
\eqref{eq:AppErro2}.
\done

\noindent{\bf Remark 1:} It can be verified that
$e_{\gamma}=\big(1+\gamma(|s|^{\frac{1}{p}}-1)\big)^{q}$. The error
factor $e_\gamma$ is increasing with $\gamma\in [0,1]$. And the
facts that $e_0=1$ and $e_1=|s|^{\frac{1}{p-1}}$ hold. So
$e_\gamma\in [1, |s|^{\frac{1}{p-1}}]$ for given $p>1$. The lemma 1
in \cite{Voice07TonDual} is a special case of Lemma 1 here with
$\gamma=1$.

\subsection{Dual Problem of the Generalized Model}
Given vectors $\Mu$ and $\Nu$, let $\bmgl=\bA^T\Mu$. The Lagrangian
of \eqref{eq:NewPrimal} is
$\sum\limits_s\big[U_s(u_s^{q})-\sum\limits_{r\in
s}(\lambda_r-\nu_s)x_r-\nu_sy_s\big]+\bc^T\Mu$. For any $s\in \mcS$,
we
\begin{equation}\label{eq:NewPrimalSub}
\begin{array}{ll}
{\rm maximize} &U_s(u_s^{q})-\sum\limits_{r\in s}(\lambda_r-\nu_s)x_r-\nu_s y_s\\[2mm]
               {\rm subject\ to }&u_s\le\gamma\sum\limits_{r\in s}x_r^{\frac{1}{q}}+(1-\gamma)y_s^{\frac{1}{q}}\\[2mm]
                                  &\bx^s\ge \bf 0,
\end{array}
\end{equation}
where $\bx^s=(x_r, r\in s)$ is the rate vector for source $s$.
We denote the optimal objective function value of
\eqref{eq:NewPrimalSub} by $W_s(\bmgl^s, \nu_s)$.
It can be verified that $W_s(\bmgl^s,\nu_s)$ is finite only if
$\lambda_r-\nu_s>0$ for all $r\in s$ and $\nu_s>0$. Otherwise,
$W_s(\bmgl^s,\nu_s)=+\infty$. In the following derivation, we assume
$\lambda_r>\nu_s>0$ for all $r\in s$ and $s\in \mcS$.

The Karush-Kuhn-Tucker condition \cite{Bert99} for
\eqref{eq:NewPrimalSub} is
\begin{IEEEeqnarray}{l}
 qU_s'(u_s^{q})u_s^{q-1}-\eta_s=0
\IEEEyessubnumber \label{eq:LtoXs}\\
\gamma (\eta_s/q)x_r^{-\frac{1}{p}}-(\lambda_r-\nu_s)=0,\quad r\in s
\IEEEyessubnumber
\label{eq:LtoXr} \\
(1-\gamma)(\eta_s/q)y_s^{-\frac{1}{p}}-\nu_s=0  \IEEEyessubnumber \label{eq:Ltols} \\
u_s  =  \gamma\sum\limits_{r\in s}x_r^{\frac{1}{q}}+(1-\gamma)y_s^{\frac{1}{q}}.
\IEEEyessubnumber \label{eq:xsbar}
\end{IEEEeqnarray}
Let $u_s^q=\bar{y}_s$, {\em  i.e.}, $u_s=\bar{y}_s^{\frac{1}{q}}$.
By \eqref{eq:LtoXs}, we have
\begin{equation}\label{eau:derive5}
\eta_s/q = U_s'(\bar{y}_s)\bar{y}_s^{\frac{1}{p}}.
\end{equation}
By (\ref{eq:LtoXr}) and \eqref{eq:Ltols}, we get
\begin{equation}\label{eau:derive6}
x_r^{\frac{1}{p}}=\frac{\gamma \eta_s/q}{\lambda_r-\nu_s},\ y_s^{\frac{1}{p}}=\frac{(1-\gamma)\eta_s/q}{\nu_s}.
\end{equation}
Substituting (\ref{eau:derive6}) into (\ref{eq:xsbar}), we obtain
$$
\bar{y}_s^{\frac{p-1}{p}}=(\eta_s/q)^{p-1}\big (\gamma^p\sum\limits_{r\in
s}(\lambda_r-\nu_s)^{1-p}+(1-\gamma)^p\nu_s^{1-p}\big).
$$
Plugging (\ref{eau:derive5}) into this equation, we get
$$
U_s'(\bar{y}_s)=\Big(\gamma^p\|\bmgl^s-\nu_s\|_{1-p}^{1-p}+(1-\gamma)^p\nu_s^{1-p}\Big)^{\frac{1}{1-p}}.
$$
Then, by the definition of the demand function
\eqref{eq:demandfunction}, we have
\begin{equation}\label{eq:NewDemand}
\bar{y}_s=D_s\Big(\big(\gamma^p\|\bmgl^s-\nu_s\|_{1-p}^{1-p}+(1-\gamma)^p\nu_s^{1-p}\big)^{\frac{1}{1-p}}\Big).
\end{equation}
By (\ref{eau:derive5}), we obtain
$(\eta_s/q)^{p}=\bar{y}_sU_s'(\bar{y}_s)^{p}$. And by
\eqref{eau:derive6}, we have that $x_r=\bar{y}_{s}\big(\frac{\gamma
U_{s}'(\bar{y}_s)}{\lambda_r-\nu_s}\big)^{p},\ \forall r\in \mcR$
and $y_s=\bar{y}_{s}(\frac{(1-\gamma)U_{s}'(\bar{y}_s)}{\nu_s})^{p},
\ \forall s\in \mcS$, where $\lambda_r=\sum_{j\in r}\mu_j$ and
$\bar{y}_s$ is defined in \eqref{eq:NewDemand}.

Finally, the Lagrangian dual problem of \eqref{eq:NewPrimal} is
\begin{equation} \label{eq:NewDual}
\Min_{\Mu\ge 0} \sum_{s\in S} W_s(\bmgl^s, \nu_s)+\bc^T\Mu.
\end{equation}
The control laws that we will propose can
be viewed as decentralized dual algorithms to solve problem \eqref{eq:NewDual} and the primal problem \eqref{eq:NewPrimal} simultaneously.

\section{A Large Family of Multi-path Dual Congestion Control Algorithms} \label{gloabl}
In this section, we will derive a large family of multi-path dual
congestion control algorithms from the generalized multi-path
utility maximization model. The stability with the absence of delay
will be stated later.
\renewcommand{\theequation}{3.\arabic{equation}}
\setcounter{equation}{0}

\subsection{Multi-path Dual Congestion Control Algorithms}
In Section \ref{utilitymodel}, a generalized utility model have been formulated, which can reduce to specific models with different parameters. We are now in a position to state a family of multi-path dual algorithm, which can be described as follows.

For all resources $j$
\begin{equation}\label{eq:RVoice_1}
\dif{\mu}_j(t)=\kappa_j(\mu_j(t))\big(z_j(t)-c_j\big)_{\mu_j(t)}^+
\end{equation}
for some positive function $\kappa_j$, and for all sources $s$,
\begin{equation}\label{eq:RVoice_2}
\dif{\nu}_s(t)=\kappa_s(\nu_s(t))\big(y_s(t)-\sum_{r\in s}
x_r(t)\big)
\end{equation}
for some positive function $\kappa_s$, where
\begin{equation}\label{eq:RVoice_3}
\begin{split}
z_j(t)&=\sum_{j\in r} x_r(t), \qquad\qquad\qquad\\
x_r(t)&=\bar{y}_{s(r)}(t)\Big(\frac{\gamma
U_{s(r)}'(\bar{y}_{s(r)}(t))}{\lambda_r(t)-\nu_{s(r)}(t)}\Big)^{p},
\end{split}
\end{equation}
\begin{equation}\label{eq:RVoice_4}
\begin{split}
     y_s(t)=& \bar{y}_{s}(t)\Big(\frac{(1-\gamma)U_{s}'(\bar{y}_s(t))}{\nu_s(t)}\Big)^{p},\\
\lambda_r(t)=&\sum_{j\in r}\mu_j(t),
\end{split}
\end{equation}
and
\begin{equation}\label{eq:RVoice_5}
\bar{y}_s=D_s\Big(\big(\gamma^p\|\bmgl^s(t)-\nu_s(t)\|_{1-p}^{1-p}+(1-\gamma)^p\nu_s(t)^{1-p}\big)^{\frac{1}{1-p}}\Big).
\end{equation}

\begin{theorem}\label{eq:EquOptimality}
Let $(\Mu,\Nu)$ be an equilibrium point of the system
\eqref{eq:RVoice_1}-\eqref{eq:RVoice_2}, and let $(\bx, \by,
\bar{\by})$ be defined through
\eqref{eq:RVoice_3}-\eqref{eq:RVoice_5}. Then $(\Mu,\Nu)$ solves the
dual problem \eqref{eq:NewDual}. And $(\bx, \by,
\bar{\by}^{\frac{1}{q}})$ is unique and solves the primal problem
\eqref{eq:NewPrimal}, where $\bar{\by}^{\frac{1}{q}}$ denotes
$(\bar{y}_s^{\frac{1}{q}}, s\in \mcS)$.
\end{theorem}

\begin{lemma}\label{eq:LiuDualDerivative}
Let $\Mu\ge \b0$ and $\bmgl=\bA^T\Mu$. Then the objective
$W(\Mu, \Nu)$ of \eqref{eq:NewDual} is differentiable with
derivative $\frac{\partial W}{\partial \mu_j}=c_j-z_j,\ \forall j\in
\mcJ, $ and $\frac{\partial W}{\partial \nu_s}=\sum_{r\in s}x_r-y_s,\
$ where $z_j=\sum_{j\in r} x_r,
x_r=\bar{y}_{s}(\frac{\gamma
U_{s}'(\bar{y}_s)}{\lambda_r-\nu_s})^{p},\lambda_r=\sum_{j\in
r}\mu_j,$
$y_s=\bar{y}_{s}(\frac{(1-\gamma)U_{s}'(\bar{y}_s)}{\nu_s})^{p}$ and
$\bar{y}_s$ is defined by \eqref{eq:NewDemand} $\forall s\in \mcS$.
\end{lemma}

\noindent {\bf Proof:}
The objective function of \eqref{eq:NewPrimal} is
{\it strictly} concave; hence, the objective function
$W(\Mu, \Nu)$ of \eqref{eq:NewDual} is convex and differentiable with $\frac{\partial W}{\partial \mu_j}=c_j-z_j,\ \forall j\in
\mcJ, $ and $\frac{\partial W}{\partial \nu_s}=\sum_{r\in s}x_r-y_s,\
\forall s\in \mcS$, where $z_j=\sum_{j\in r} x_r$ and $(\bx, \by)$ solve problem
\eqref{eq:NewPrimalSub} \cite{Bert99}. By the definition of the system \eqref{eq:RVoice_1}-\eqref{eq:RVoice_5}, we get the desired results.
\done

\subsection{Global Stability}

Assume that the matrix $\bA$ has full row rank and
$u_s^{q-1}U'_s(u_s^q)$ is strictly decreasing. This condition and Assumption H are
sufficient to deduce that the system
\eqref{eq:RVoice_1}-\eqref{eq:RVoice_5} has a unique equilibrium
point. The first is the general assumption to make sure the unique
equilibrium point when analyzing the dual congestion control
algorithms, such as \cite{Paganini02}, \cite{Kelly03} and
\cite{Voice07TonDual}. Note, the assumption is needed only for the
links that would be a bottleneck; so our assumption is quite
generic. Following is the first main result of this paper with a strict proof, where the proof is completely different from the heuristic one given in \cite{Voice07TonDual}.

\begin{theorem}\label{theorem:GlobalS}
Given the system defined by \eqref{eq:RVoice_1}-\eqref{eq:RVoice_5}.
Then the unique equilibrium point $(\Mu^*,\Nu^*)$ is globally
asymptotically stable.
\end{theorem}
The proof of Theorem \ref{theorem:GlobalS} can be found in Appendix.

\section{Delay Stability} \label{local}
As transmission delay universally exists in network environment, in this section we will present a particular scheme, where the
proposed algorithms can achieve stability in the presence of propagation delays.
\renewcommand{\theequation}{4.\arabic{equation}}
\setcounter{equation}{0}

\subsection{Choice of Scheme}
When we include propagation delays, we get the following algorithms,
for which we can provide scalable, decentralized stability
conditions.

For all resources $j$,
\begin{equation}\label{eq:RVoiceD_1}
\dif \mu_j(t)=\frac{\kappa_j\mu_j(t)}{p}\big(\sum_{j\in r}
x_r(t-T_{rj})-c_j\big)_{\mu_j(t)}^+
\end{equation}
for some positive constant $\kappa_j$, and for all sources $s$,
\begin{equation}\label{eq:RVoiceD_2}
\dif \nu_s(t)=\frac{\kappa_s\nu_s(t)}{p}\big(y_s(t)-\sum_{r\in s}
x_r(t-T_r)\big),
\end{equation}
\begin{equation}\label{eq:RVoiceD_3}
\begin{aligned}
\dif \bar{y}_s(t)=\frac{q\rho_s}{p}\left({\gamma\sum_{r:r\in s}x_r(t-T_r)^{\frac{1}{q}}}\right.\\
\left.{+(1-\gamma)y_s(t)^{\frac{1}{q}}-\bar{y}_s(t)^{\frac{1}{q}}}\right)
\end{aligned}
\end{equation}
for some positive constants $\kappa_s$ and $\rho_s$, where
\begin{equation}\label{eq:RVoiceD_4a}
x_r(t)=\bar{y}_{s(r)}(t)\Big(\frac{\gamma U_{s(r)}'(\bar{y}_{s(r)}(t))}{\lambda_r(t)-\nu_{s(r)}(t)}\Big)^{p},
\end{equation}
\begin{equation}\label{eq:RVoiceD_4b}
y_s(t)= \bar{y}_{s}(t)\Big(\frac{(1-\gamma)U_{s}'(\bar{y}_s(t))}{\nu_s(t)}\Big)^{p},
\end{equation}
and
\begin{equation}\label{eq:RVoiceD_5}
\begin{split}
\lambda_r(t)=&\sum_{j\in r}\mu_j(t-T_{jr}).
\end{split}
\end{equation}
Compared with \eqref{eq:RVoice_1}-\eqref{eq:RVoice_5}, we set
$\kappa_j\mu_j(t)/p$ and
$\kappa_s\nu_s(t)/p$ as the dynamic gain factor for resource
$j$ and source $s$ respectively. In addition, we relax the algebraic
equation \eqref{eq:RVoice_5} to differential equation
\eqref{eq:RVoiceD_3} for the existence of delay.

We first give some properties of the equilibrium point, which are
useful in the proof of the main result.

\begin{lemma}\label{le:RVLocalS1}
We have
$a_s:=-\frac{U''_s(\bar{y}_s)}{U'_s(\bar{y}_s)}-\frac{1}{p\bar{y}_s}>0$,
where $\bar{y}_s=u_s^q$.
\end{lemma}
\noindent {\bf Proof:}
We have $b_s:=(u_s^{q-1}U_s(u_s^q))'<0$ for we suppose Assumption H holds. It can be checked that
$$\begin{array}{ll}
b_s&=(q-1)u_s^{q-2}U'_s(\bar{y}_s)+qu_s^{2(q-1)}U^{''}_s(\bar{y}_s)\\
   &=-qu_s^{2(q-1)}U'_s(\bar{y}_s)a_s.
\end{array}$$
Combining it with $U'_s(\bar{y}_s)$ nonnegative by $U_s(\bar{y}_s)$ concave with $\bar{y}_s$, we get the desired result.
\done

\begin{lemma}\label{le:RVLocalS2}
Let $(\Mu, \Nu, \bar{\by})$ be an equilibrium point of system
\eqref{eq:RVoiceD_1}-\eqref{eq:RVoiceD_3} and $(\bx, \by, \bmgl)$
defined by \eqref{eq:RVoiceD_4a}-\eqref{eq:RVoiceD_5}. Then for each $r\in \mcR$ we have
that
\begin{equation}\label{eq:RVLocalS21}
\frac{x_r}{\bar{y}_{s(r)}^{\frac{1}{p}}U'_{s(r)}}=\frac{\gamma
x_r^{\frac{1}{q}}}{\lambda_r-\nu_{s(r)}}
\end{equation}
and
\begin{equation}\label{eq:RVLocalS22}
1+\frac{\nu_{s(r)}}{\lambda_r-\nu_{s(r)}} + \sum_{j\in r}\frac{\mu_j}{\lambda_r-\nu_{s(r)}}\le \frac{2}{\gamma}.
\end{equation}
\end{lemma}

\noindent {\bf Proof:}
Firstly, we have $x_r=\bar{y}_{s(r)}\left(\frac{\gamma U'_{s(r)}(\bar{y}_{s(r)})}{\lambda_r-\nu_{s(r)}}\right)^p$
 by \eqref{eq:RVoiceD_4a}. Then power the bothside of this equation with $1/p$, we get
\begin{equation} \label{eq:lemma4_proof1}
 x_r^{\frac{1}{p}}=\bar{y}_{s(r)}^{\frac{1}{p}}\frac{\gamma U'_{s(r)}(\bar{y}_{s(r)})}{\lambda_r-\nu_{s(r)}}.
\end{equation}
Multiply \eqref{eq:lemma4_proof1} with $x_r^{1/q}$ and combine the fact $1/p+1/q=1$, we have the equation \eqref{eq:RVLocalS21}.

Similarly with \eqref{eq:lemma4_proof1}, we have
\begin{equation}\label{eq:lemma4_proof2}
y_{s(r)}^{\frac{1}{p}}=\bar{y}_{s(r)}^{\frac{1}{p}}\frac{(1-\gamma) U'_{s(r)}(\bar{y}_{s(r)})}{\nu_{s(r)}}
\end{equation}
from \eqref{eq:RVoiceD_4b}. Combine \eqref{eq:lemma4_proof1} with \eqref{eq:lemma4_proof2}, we get
$$\frac{\nu_{s(r)}}{\lambda_r-\nu_{s(r)}}=(\frac{1}{\gamma}-1)\left(\frac{x_r}{y_{s(r)}}\right)^{\frac{1}{p}}.$$
Combining it with $x_r\le y_s$, we have $\frac{\nu_{s(r)}}{\lambda_r-\nu_{s(r)}}\le \frac{1}{\gamma}-1$.
Then
$$ \begin{array}{ll}
    & 1+\frac{\nu_{s(r)}}{\lambda_r-\nu_{s(r)}} + \sum_{j:j\in r}\frac{\mu_j}{\lambda_r-\nu_{s(r)}}\\
=   & 1+\frac{\nu_{s(r)}}{\lambda_r-\nu_{s(r)}} + \frac{\lambda_r}{\lambda_r-\nu_{s(r)}}\\
=   & 2+\frac{2\nu_{s(r)}}{\lambda_r-\nu_{s(r)}}\\
\le &\frac{2}{\gamma}.
\end{array}$$
\done

\subsection{Main Result}
We now turn to our main concern, the local stability of the system \eqref{eq:RVoiceD_1}-\eqref{eq:RVoiceD_5}.

Define a link $j$ to be {\em almost saturated} if at which both $\mu_j=0$ and condition
\begin{equation}\label{eq:strictC}
\sum_{r:j\in r} x_r =c_j, j\in \mcJ
\end{equation}
holds. We thus rule out the possible degeneracy that {\rm both} terms in the product \eqref{eq:RVoiceD_1} might vanish.

\begin{theorem}\label{th:RVoiceLocalS}
Let $(\Mu, \Nu, \bar{\by})$ be an equilibrium point of system
\eqref{eq:RVoiceD_1}-\eqref{eq:RVoiceD_5} with no almost saturated links, and suppose that for all $j\in \mcJ$,
\begin{equation}\label{eq:MVlocalstability1}
\kappa_j\sum_{j\in r}x_rT_r<\gamma\frac{\pi}{4},
\end{equation}
and for all $s\in \mcS$,
\begin{equation}\label{eq:MVlocalstability2}
\kappa_s\sum_{r\in s}x_rT_r<\gamma\frac{\pi}{4}
\end{equation}
and
\begin{equation}\label{eq:MVlocalstability3}
\rho_sa_s\sum_{r\in s}x_r^{\frac{1}{q}}T_r<\frac{\pi}{4}.
\end{equation}
Then there exists a neighborhood $\mcN$ of $\Mu$
such that for any initial trajectory $((\Mu(t), \Nu(t),
\bar{\by}(t)), t\in (-T_{\max},0))$ with $\Mu(t)$ lying within the neighborhood
$\mcN$, $(\Mu(t),\Nu(t))$ converge as $t\to \infty$ to the solution
$(\Mu, \Nu)$ to the optimization problem \eqref{eq:NewDual} and
$(\bx(t),\by(t),\bar{\by}(t)^{\frac{1}{q}})$ converge as $t\to
\infty$ to the solution $(\bx, \by, \bar{\by}^{\frac{1}{q}})$ to the
optimization problem \eqref{eq:NewPrimal}, where $T_{\max}=\max(T_r, r\in \mcR)$.
\end{theorem}

The proof of Theorem \ref{th:RVoiceLocalS} can be found in Appendix.

\subsection{Result for $\alpha$-fair Utility Function}
Corresponding to weighted $\alpha$-fair utility function, defined by
equation \eqref{eq:utilityfunction}, the $a_s$ defined in Lemma \ref{le:RVLocalS1} becomes
$\frac{\alpha p-1}{p\bar{y}_s}$. Then the local stability
condition \eqref{eq:MVlocalstability3} reduces to
$$\rho_s(\alpha p-1)\frac{\sum_{r\in
s}x_r^{\frac{1}{q}}T_r}{\bar{y}_s}<p\frac{\pi}{4}.$$
Since $\bar{y}_s\ge y_s=\sum_{r:r\in s}x_r$ and
$x_r^{\frac{1}{q}}<x_r$, the conditions \eqref{eq:MVlocalstability1}, \eqref{eq:MVlocalstability2} and
\begin{equation}\label{eq:MVlocalstabilityS3}
\rho_s(\alpha p-1)\frac{\sum_{r\in s}x_rT_r}{y_s}<
p\frac{\pi}{4}
\end{equation}
are sufficient to derive the local stability for this special case.

These conditions are attractive because they are local and
decentralized. Let the maximize available rate for source $s$ be
$M_s$. They lead to a highly scalable parameter choice scheme: each
source and link chooses their gain parameters to be
$\kappa_s=\frac{\gamma\kappa}{M_s\bar{T}_s},
\rho_s=\frac{p\kappa}{(\alpha p -1)\bar{T}_s}$ and
$\kappa_j=\frac{\gamma\kappa}{c_j\bar{T}_j}$ for some $\kappa\in
(0,\frac{\pi}{4})$, where $\bar{T}_s=\frac{\sum_{r:r\in
s}x_rT_r}{y_s}$ is the average round trip time of packets
transmitted by source $s$ and $\bar{T}_j=\frac{\sum_{r:j\in
r}x_rT_r}{\sum_{r: j\in r}x_r}$ is the average round trip time of
packets passing through resource $j$. As a desirable feature, the
gain parameters in the proposed algorithms can be derived from local
information only. Independence of state information in networks
leads conditions for delay stability to be scalable and
decentralized.

\noindent{\bf Remark 2:} If $\gamma=1$, the conditions
\eqref{eq:MVlocalstability1} and \eqref{eq:MVlocalstabilityS3} reduce to
the sufficient ones determined by Voice \cite{Voice07TonDual}
except they use an estimation of the average round trip time of
packets transmitted by source $s$. The delay stability condition for
single-path fair dual algorithm given in \cite{Kelly03} has minor
difference from these condition, which was derived by linearizing
the system about the flow rate $x_r(t)$.

\section{Simulation Experiments}\label{simu}
We will further investigate the proposed algorithm by simulation
experiments, which mainly focus on two aspects:

a) To evaluate the performance of the proposed algorithm in terms of
stability and convergence rate. We implement our algorithm in Matlab and explore the influence of
parameters on these two features.

b) To study the performance of the proposed algorithm in network
environment. We use NS2 to implement the proposed algorithm in a
window-based network environment and make performance comparison
with the other existing two algorithms.

In the sequel, we take the utility with the form of
\eqref{eq:utilityfunction} with $\alpha=1$ and $w_s=1$ for all $s$.

\subsection{Matlab Simulation}
\begin{figure}[t!]
\centering
\includegraphics[width=7cm,height=3.5cm]{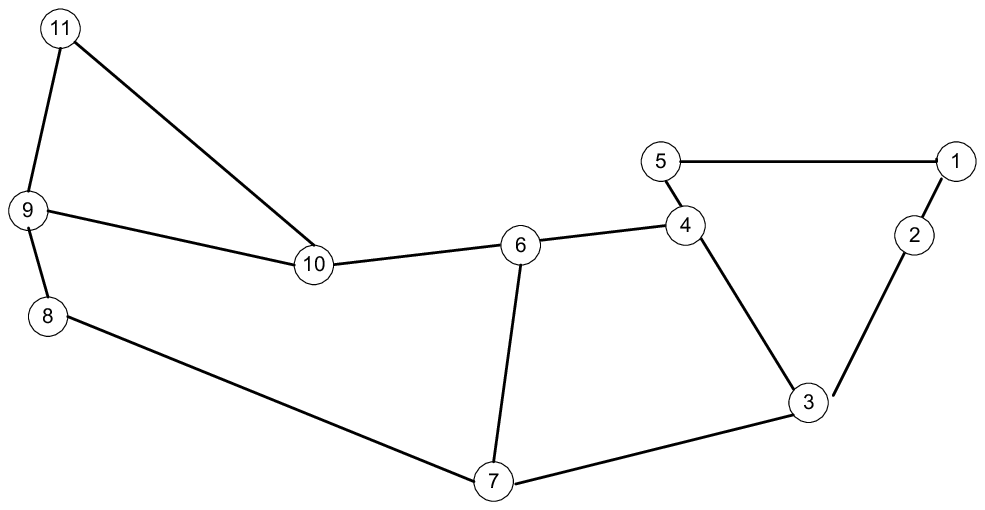}\\
\scriptsize{(a) Used in Matlab simulation}\\
\centering
\includegraphics[width=5cm,height=2cm]{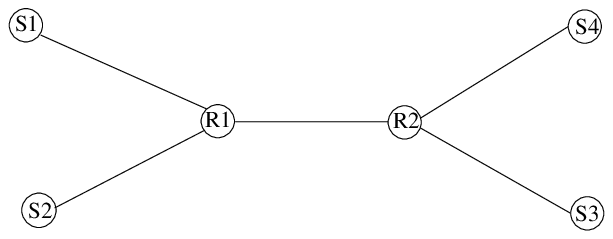}\\
\scriptsize{(b) Used in NS2 simulation}\\
\renewcommand{\figurename}{Fig.}
\caption{Network Topology }
\label{fig:Topo}
\end{figure}

To really achieve the best tradeoff between stability and convergence one has to carefully select a few parameters.
To make our work really usefully for network designers and administers, we provide some discussion and rules of thumb to select the values.
We use Matlab to implement the proposed algorithm and investigate
how parameters affect the algorithm's performance. The network
topology used is Abilene backbone network, shown in
Fig.\ref{fig:Topo} (a), which have fourteen 100-Mb/s links with 2-ms
delay. The discussion explain the intuitive meaning of changing a parameter in each direction.

First, we select a set of $\gamma$ with fixed values, and vary the
gains to explore the trade-off between the convergence rate and
stability; Then, we vary the values of parameter $\gamma$ within the
interval $[0,1]$, and study how to tune the gains to make the system
achieve the optimal performance.

\noindent{\bf Stability and  convergence rate.} First, we set $p=2, \gamma=0.2$ and observe how the gains
$\kappa_j$ and $\kappa_s$ to trade off between the
convergence rate and the stability.

\begin{figure}[t!]
\centering
\includegraphics[width=6cm,height=4cm]{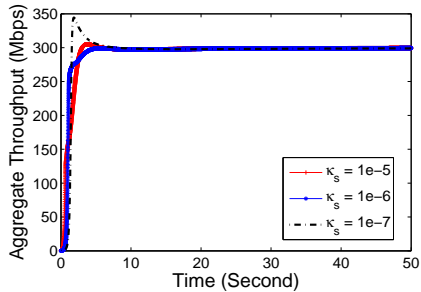}\\
\scriptsize{(a) For three different $\kappa_s$}\\
\centering
\includegraphics[width=6cm,height=4cm]{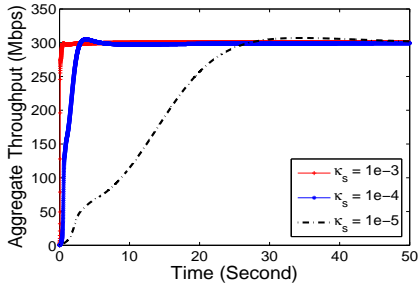}\\
\scriptsize{(b) For three different $\kappa_j$}\\
\renewcommand{\figurename}{Fig.}
\caption{Aggregate throughput of all routes vs. gains for the network
topology shown in Fig.\ref{fig:Topo}(a)} \label{fig:varyGains}
\end{figure}

For simplicity, we first fix link gain $\kappa_j$ to be $10^{-4}$,
and study the effect of source gain $\kappa_s$ on system
performance. Here we run the simulation with three different
$\kappa_s$ values $10^{-5}$, $10^{-6}$ and $10^{-7}$ respectively.
In each simulation we set step to 5ms and run the simulation for
50s. The results is shown in Fig.\ref{fig:varyGains} (a), where axis
$Y$ represents the aggregate throughput of four routes. All
simulations with three different parameters will achieve global
stability roughly at 5s. We notice that, however, when $\kappa_s=
10^{-6}$, the coverage is slightly slow. And when $\kappa_s= 10^{-7}$,
there exists big oscillation before stability reaches. It seems that
$10^{-5}$ is the best choice among these three values of $\kappa_s$.

And then we fix source gain $\kappa_s$ to be $10^{-5}$ to explore
the influence of link gain $\kappa_j$ on system performance. We run
simulations with three different $\kappa_j$ values $10^{-3}$,
$10^{-4}$ and $10^{-5}$ respectively. Each simulation runs 50s with
step of 5ms. The simulation results are illustrated in
Fig.\ref{fig:varyGains} (b), where the trends of the aggregate throughput
are similar to those in Fig.\ref{fig:varyGains} (a).  Simulation with
a larger value of link gain converges faster and meanwhile
experiences more severe oscillation. We notice that, with
$\kappa_j=10^{-4}$ and $\kappa_s=10^{-5}$, the algorithm can achieve
a better trade-off between the stability and convergence rate.

\begin{figure}[t!]
\centering
\includegraphics[width=6cm,height=4cm]{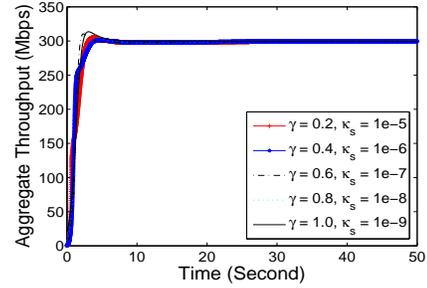}\\
\scriptsize{(a) Aggregate throughput of all routes for different $\gamma$ and
$\kappa_s$}\\
\centering
\includegraphics[width=6cm,height=4cm]{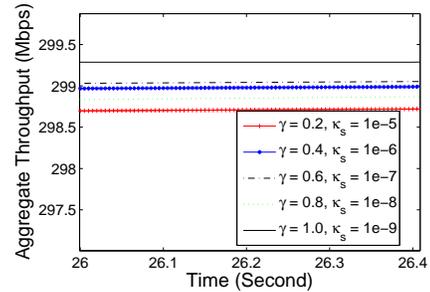}\\
\scriptsize{(b) Optimal throughput of all routes for different $\gamma$ and
$\kappa_s$}\\
\renewcommand{\figurename}{Fig.}
\caption{Algorithm's performance with different $\gamma$ for the network topology shown in Fig.\ref{fig:Topo}(a)}
\label{fig:varyGamma}
\end{figure}

\begin{figure*}[t!]
\begin{minipage}[t]{6cm}
\centering
\includegraphics[width=6cm,height=4cm]{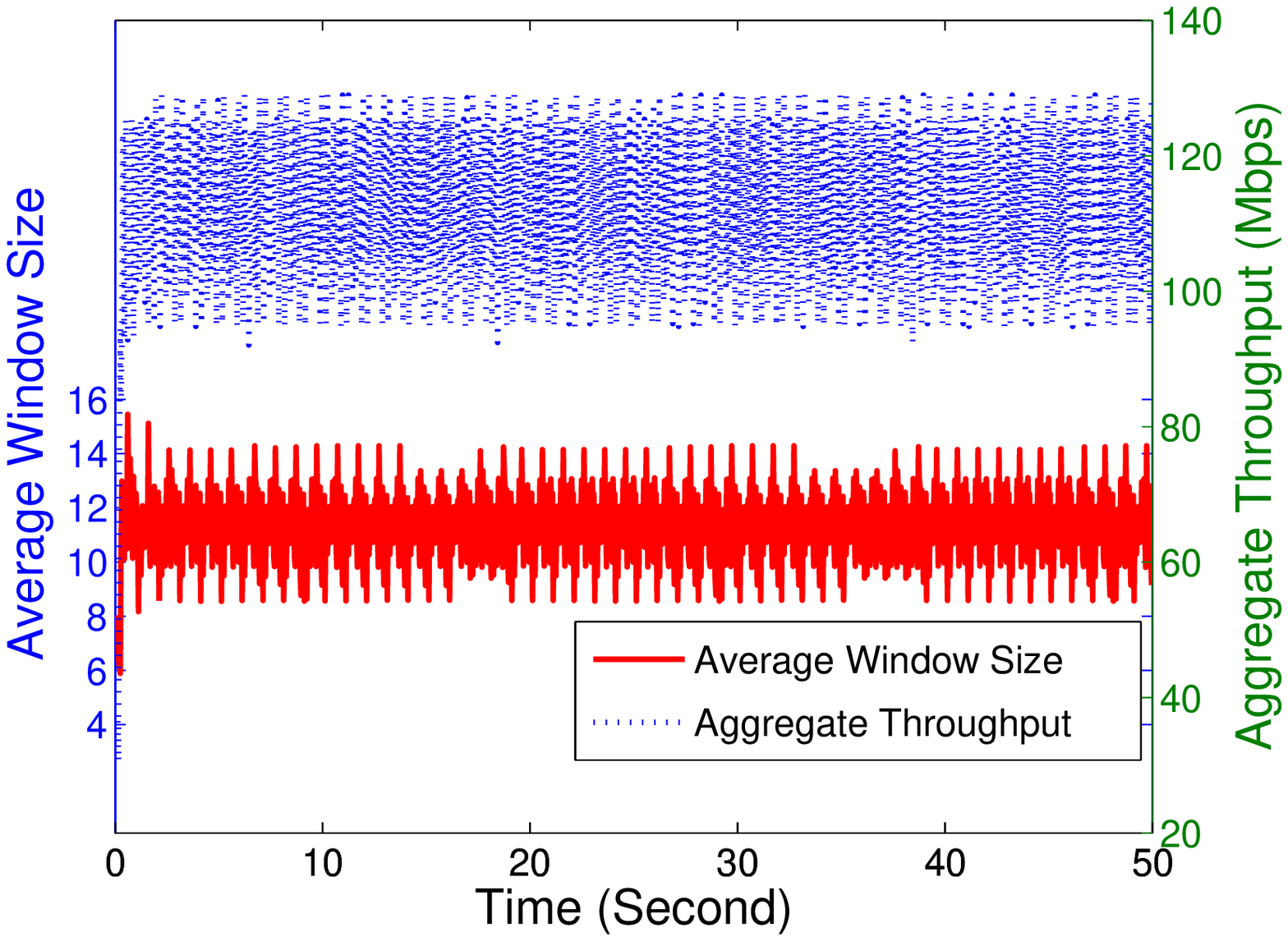}\\
\scriptsize{(a) With Algorithm in \cite{KellyVoice05}}\\
\end{minipage}
\ \
\begin{minipage}[t]{6cm}
\centering
\includegraphics[width=6cm,height=4cm]{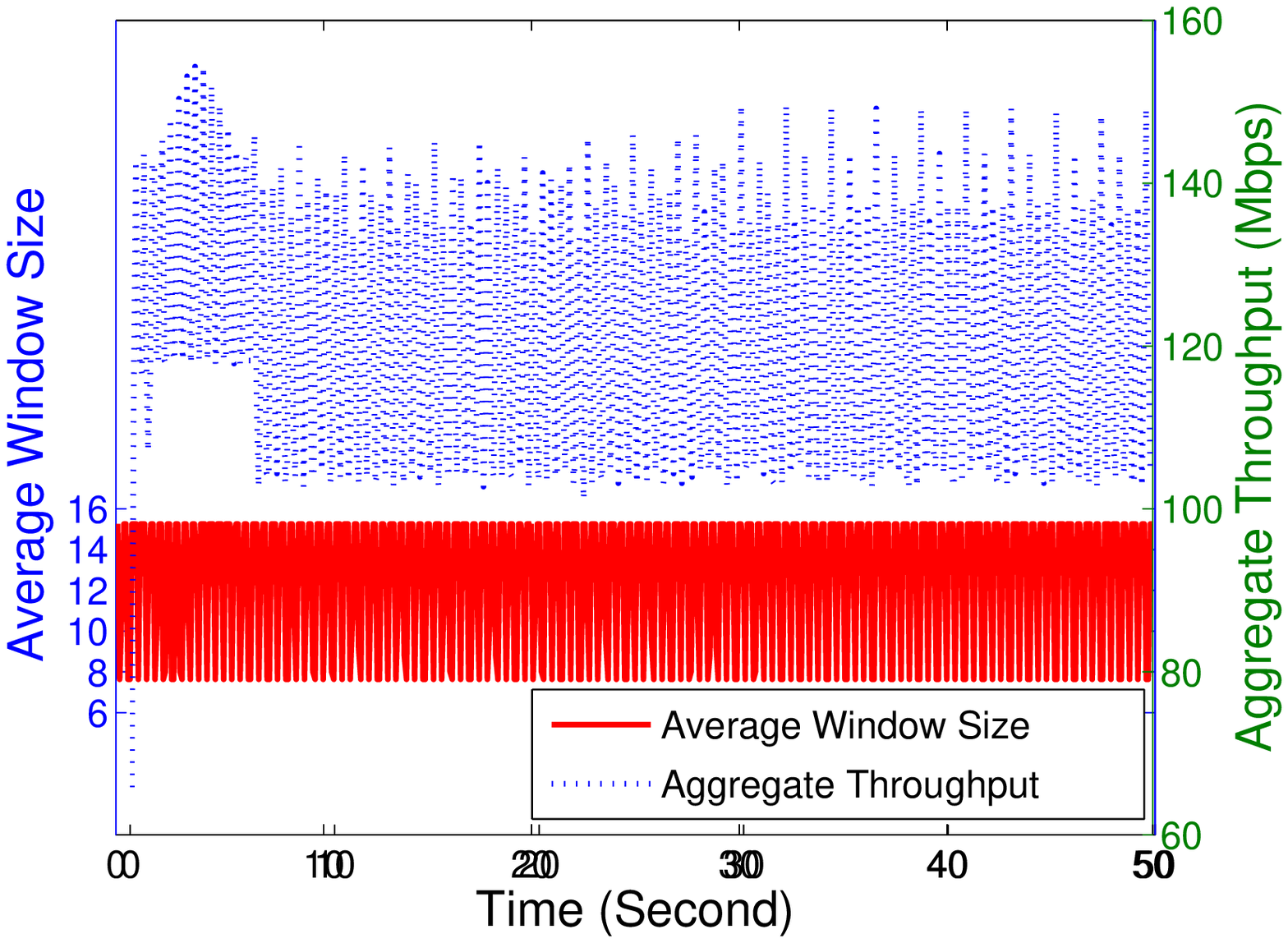}\\
\scriptsize{(b) With Algorithm in \cite{Voice07TonDual}}\\
\end{minipage}
\ \
\begin{minipage}[t]{6cm}
\centering
\includegraphics[width=6cm,height=4cm]{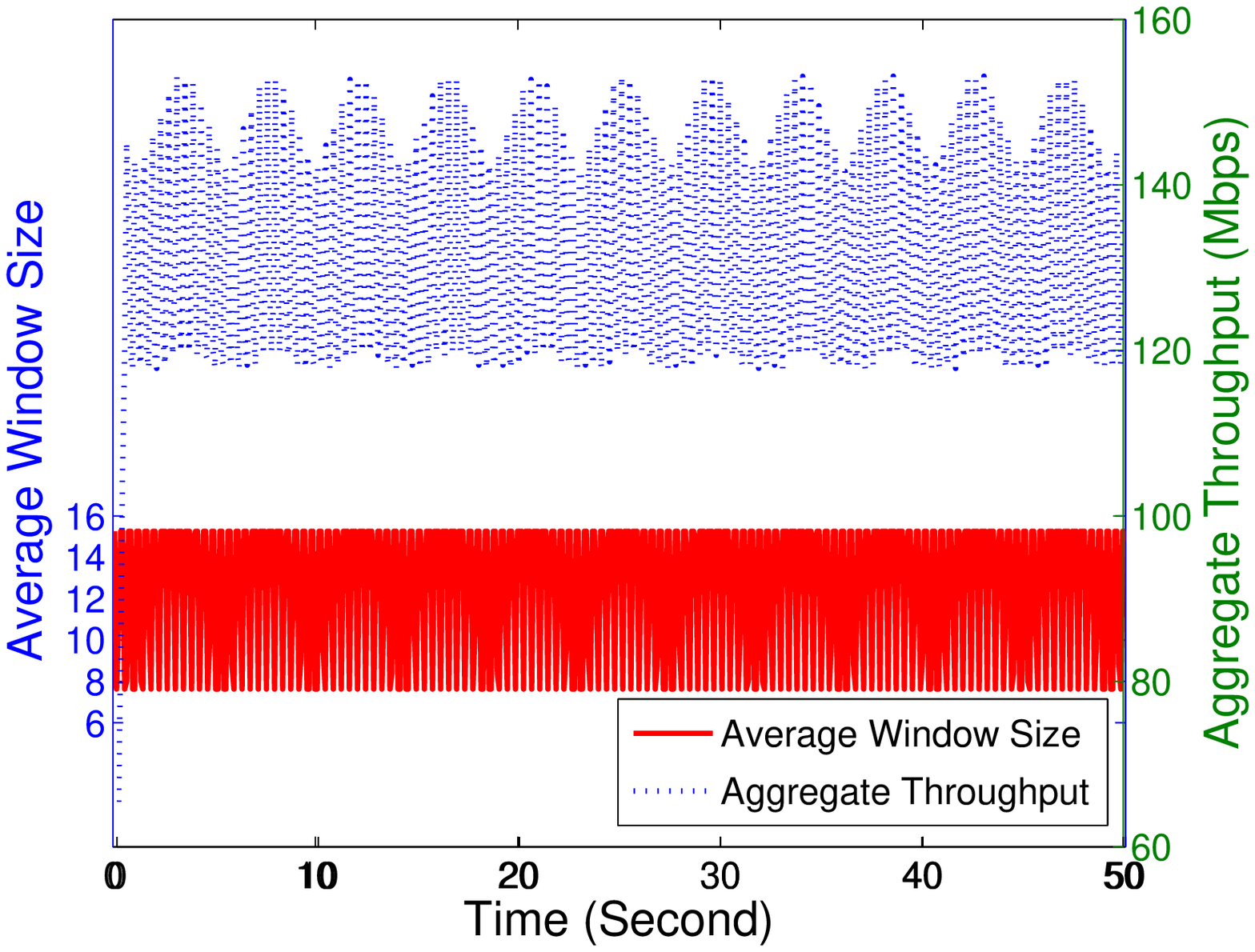}\\
\scriptsize{(c)  With the proposed algorithm}\\
\end{minipage}
\renewcommand{\figurename}{Fig.}
\caption{Average window size and throughput for the network topology
shown in Fig.\ref{fig:Topo} (b)} \label{fig:NSresults}
\end{figure*}

\noindent{\bf Approximation Parameters.} In the proposed algorithms, there exist two parameters, $p$ and $\gamma$ which control the error factor $e_\gamma$ in Lemma 1.
As $p$ also exists in the algorithm proposed by Voice
\cite{Voice07TonDual}, here we only consider the influence of
$\gamma$ on system performance.

As shown in Fig.\ref{fig:varyGains} (b), simulation with link gain
$\kappa_j=10^{-4}$ converges fast, we thus set $p=2$ and
$\kappa_j=10^{-4}$ in the following simulations. We run simulations
with $\gamma$ varying from 0 to 1, and explore how source gain
$\kappa_s$ should react to different values of $\gamma$ in order to
achieve the optimal performance. Each simulation runs with step =
5ms and lasts for 50s.

The simulation results exhibit in Fig. \ref{fig:varyGamma}. From
Fig.\ref{fig:varyGamma} (a), we notice that parameter $\gamma$ will
influence the characteristics of both convergence and oscillation.
Simulation with a smaller value of $\gamma$ experiences greater
oscillation before achieving stability. The value of $\kappa_s$
should decrease with the increase of $\gamma$ to guarantee the
simulation converges. Besides, the convergence throughput, at which
the algorithm achieve global stability, is influenced by the choice
of $\gamma$. Fig.\ref{fig:varyGamma} (b) shows that a smaller value
of $\gamma$ will lead to a smaller gap between the optimal
throughput and the one obtained with the proposed algorithm.

\subsection{NS2 Simulation}
The proposed congestion control algorithm is derived from a rate-based model.
TCP, however, is a window-based control protocol. Now we use NS2 to implement
TCP-Reno for further investigation. 

The topology used is shown in Fig.1 (b). Here we consider the case of three source-destination pairs, namely from 1 to 4, 
1 to 2 and 1 to 3, and each link-capacity along the route is set to 100Mb/s with a one way propagation delay of 2ms. In simulation, each source-destination pair is attached to multiple routes. For example, pair 1 and 2 has two routes, from 1 to 2 directly and from 1 to 2 via 3.

1)Stability and convergence rate
According to the rate-based simulation, we implement the proposed algorithm with the parameters shown in Table II, which guarantee the algorithm achieving its optimal performance. The simulation lasts for 5 seconds and 7 TCP connections start at the same time and during the whole simulation.

The performance of the proposed algorithm in terms of each source-destination's aggregated throughput and the whole network's aggregated throughput is shown in Fig.4. As is shown, the aggregated throughput with the proposed algorithm achieves equilibrium at about 1.5s.

2)Performance comparison with different values of $\gamma$
Now we vary the values of $\gamma$ within the interval [0,1] and achieve the optimal performance. Throughput of the three source-destination pairs is 
plotted in Fig.5 (a)~(e), where $\gamma$ is set to 0.2, 0.4, 0.6, 0.8, 1.0 respectively.

All the three source-destination pairs can achieve equilibrium after a period of oscillation, however, with different approximate aggregated throughput
when $\gamma$ varies from 0.2 to 1.0. According to Fig.6, better consideration of fairness is shown when the equilibrium conditions are achieved
if $\gamma \rightarrow 0$ and when $\gamma \rightarrow 1$, there exists a trade-off among different source-destination pairs in terms of aggregated throughput.
Thus, possibly, we can set $\gamma$ to different values to meet different demands when facing different situations of the network, meanwhile remaining equilibrium conditions.

\begin{table}[!t]
\renewcommand{\arraystretch}{1.3}
\caption{Parameters in NS2 simulations}
\label{tab:NSpara} \centering
\begin{tabular}{||c|c|c||}
\hline Algorithm                      &  Parameters              & Gains \\
\hline Kelly \cite{KellyVoice05}      &   $\beta_j=1$            & $\kappa_r=10^{-3}$ \\
\hline Voice \cite{Voice07TonDual}    &   $p=2$ &$\kappa_j=10^{-4}$,
$\kappa_s=10^{-5}$ \\ \hline Liu &   $p=2,\gamma=0.4$       & $\kappa_j=10^{-4}, \kappa_s=10^{-5}$\\
\hline
\end{tabular}
\end{table}

Finally, throughput for TCP connections does not make sense. Particularly when experimenting with congestion control.
It counts even re-transmits as useful packets. We measure the goodput. Usually goodput can be substantially different than throughput when the congestion control makes the window oscillate. We would like to know, how our schemes with their oscillations cpmpare with the alternatives.

\section{Conclusion}\label{conclusion}
This paper considers the well known problem of joint multi-path routing and flow control, and brings new insight to congestion control algorithms. Specifically, we propose a generalized multi-path utility
maximization model, and then derive a family of muti-path dual
congestion control algorithm. Based on the results in this paper,
one can understand the unstability of the natural muti-path dual
congestion control algorithm which is a special cases in the
proposed family. The one proposed in \cite{Voice07TonDual}, a
special case in this family, is at a risk to choose a sufficient
large $p$ to approximation the solution of the original problem.
Simulation results show that the proposed algorithm can achieve a
more optimal and stable aggregate throughput with an appropriate
value of the average window size than the others multi-path
congestion control algorithms.

Future work is mainly focused on enriching simulation experiments of the proposed dual multi-path congestion control algorithms.
The excellent performance of the proposed algorithms in real network
environment is the ultimate goal we are pursuing.


\section{Appendix}
The proof of Theorem \ref{theorem:GlobalS} is presented as follows:

\noindent {\bf Proof:} The proof is based on Lasalle's invariance principle applied to a
suitable Lyapunov function. Now we introduce the candidate Lyapunov
function $V(\Mu,\Nu)=W(\Mu,\Nu).$ For any state vector
$(\Mu(t),\Nu(t))\neq (\Mu^*,\Nu^*)$ of system
\eqref{eq:RVoice_1}-\eqref{eq:RVoice_2}, it can be seen that
$(\Mu(t),\Nu(t))$ is feasible for the dual problem
\eqref{eq:NewDual}. For $(\Mu^*,\Nu^*)$ is the unique solution of
\eqref{eq:NewDual}, we have $W(\Mu,\Nu)>W(\Mu^*,\Nu^*)$.

We now take the derivative of $V(\Mu,\Nu)$ along trajectories of our system:
$$\begin{array}{rl}
\dif{V}=&\sum_{j\in J}\frac{\partial W}{\partial \mu_j}\dot{\mu}_j+\sum_{s\in S}\frac{\partial W}{\partial \nu_s}\dot{\nu}_s\\
       =&\sum_{j\in J}(c_j-z_j)\dot{\mu}_j+\sum_{s\in S}(\sum_{r\in s}x_r-y_s)\dot{\nu}_s\\
       =&\sum_{j\in J}v_j-\sum_{s\in \mcS} \kappa_s(\nu_s)(y_s-\sum_{r\in s}x_r)^2,
\end{array}$$
where we have denoted $v_j:=(c_j-z_j)\dot{\mu}_j$. Note that the second equality follows from Lemma 2. We will now show
that $v_j\le 0$ for each $j$. For this, we must apply the dynamic
equations \eqref{eq:RVoice_1}-\eqref{eq:RVoice_2}, and distinguish between the two
cases:

(a) $\mu_j>0$. Here $v_j=-\kappa_j(\mu_j)(z_j-c_j)^2.$

(b) $\mu_j=0$. Here $v_j=\kappa_j(\mu_j)(c_j-z_j)\max(0,z_j-c_j)$.
There are two cases:
$$\begin{array}{l}
v_j=0 \quad {\rm for }\  z_j<c_j,\\
v_j=-\kappa_j(\mu_j)(z_j-c_j)^2 \quad {\rm for}\ z_j\ge c_j.
\end{array}$$

We thus confirm that $v_j\le 0$ for every $j$, and thus $\dot{V}\le
0$. Invoking Lyapunov's stability theorem, we conclude that the
trajectory $(\Mu(t),\Nu(t))$ must remain bounded over time, and that
the equilibrium point $(\Mu^*,\Nu^*)$ is stable in the sense of
Lyapunov: trajectories starting close to it will remain inside a
neighborhood.

To establish the stronger claim of {\em asymptotic} stability, we
must show that trajectories will converge to equilibrium as time
goes to infinity. We do this by means of Lasalle's invariance
principle (see, e.g. \cite{LasalleP}). To apply it, we must study
the set of states $(\Mu,\Nu)$ where the Lyapunov derivative is zero,
or equivalently $v_j=0$ for each $j$ and $y_s=\sum_{r\in s}x_r$ for
each $s$. Reviewing the cases above, we find that $v_j=0$ can only
happen when either
\begin{enumerate}
\item [(i)] $z_j=c_j$, or
\item [(ii)] $z_j<c_j$ and $\mu_j=0$.
\end{enumerate}

The Lasalle principle is based on identifying an {\em  invariant}
set inside the set $\{(\Mu,\Nu): \dot{V}=0\}$. For this purpose,
suppose a trajectory $\Mu(t)$ moves inside this set. Then for each
$j$ we must have $\mu_j(t)=\mu^0_j$, where $\Mu^0$ is the initial
state.

To see this, first note that if $\mu^0_j=0$ for a certain $j$, then
it must remain this way because $z_j-c_j\le 0$ under both
alternative (i) and (ii). Using this fact again, now
\eqref{eq:RVoice_1} implies that $\dot{\mu}_j=0$ under both
alternatives, so $\mu_j(t)=\mu^0_j$.

If instead $\mu^0_j>0$, we are initially in alternative (i) and thus
$\mu_j$ stays constant $\mu_j(t)=\mu^0_j$ due to
\eqref{eq:RVoice_1}. Then we stay in this alternative indefinitely.

Now we observe that for a trajectory satisfying the alternatives (i) or (ii) is the unique equilibrium $(\Mu^*,\Nu^*)$.
\done

The proof of Theorem \ref{th:RVoiceLocalS} is presented as follows:

\noindent {\bf Proof:} Initially assume that $\mu_j>0$ for $j\in \mcJ$, and thus \eqref{eq:strictC} holds
for each $j\in \mcJ$. Later we shall see that the assumption is without loss of generality.

Let $x_r(t)=x_r+u_r(t), y_s(t)=y_s+v_s(t)$,
$\bar{y}_s(t)=\bar{y}_s+\bar{v}_s(t), \lambda_r(t)=\lambda_r+v_r(t),
\mu_j(t)=\mu_j+w_j(t), \nu_s(t)=\nu_s+w_s(t).$ Then, linearizing the
system \eqref{eq:RVoiceD_1}-\eqref{eq:RVoiceD_3} about $\Mu, \Nu$
and $\bar{\by}$, and using the relation \eqref{eq:RVLocalS21}, we
obtain the following equations
\begin{equation}\label{eq:RVLocalS1}
\dif w_j(t)=\frac{\kappa_j\mu_j}{p}\sum_{j\in r}u_r(t-T_{rj}),
\end{equation}
\begin{equation}\label{eq:RVLocalS2}
\dif w_s(t)=\frac{\kappa_s\nu_s}{p}(v_s(t)-\sum_{r\in
s}u_r(t-T_{r})),
\end{equation}
\begin{equation}\label{eq:RVLocalS3}
\begin{aligned}
\dif \bar{v}_s(t)=\frac{\rho_s}{p}\left({\gamma\sum_{r\in
s}x_r^{-\frac{1}{p}}u_r(t-T_r)}\right.\\
\left.{+(1-\gamma)y_s^{-\frac{1}{p}}v_s(t)-\bar{y}_s^{-\frac{1}{p}}\bar{v}_s(t)}\right),
\end{aligned}
\end{equation}
where
$$u_r(t)=-px_r\left(a_s\bar{v}_s(t)+\frac{\sum_{j\in r}w_j(t-T_{jr})}{\lambda_r-\nu_s}-\frac{w_s(t)}{\lambda_r-\nu_s}\right),$$
$$v_s(t)=-py_s\left(a_s\bar{v}_s(t)+\frac{w_s(t)}{\nu_s}\right),$$
and $a_s=-\frac{U''_s}{U'_s}-\frac{1}{p\bar{y}_s}>0$ by Lemma
\ref{le:RVLocalS1}.

Let us overload notation and write $u_r(\omega), v_s(\omega),
\bar{v}_s(\omega)$ and $v_r(\omega), w_j(\omega), w_s(\omega)$ as the
Laplace transforms of $u_r(t), v_s(t),\bar{v}_s(t)$ and $v_r(t),
w_j(t), w_s(t)$, respectively. We may deduce from
\eqref{eq:RVLocalS1}-\eqref{eq:RVLocalS3},
$$\omega w_j(\omega)=\frac{\kappa_j\mu_j}{p}\sum_{j\in r}e^{-\omega
T_{rj}}u_r(\omega),$$
$$\omega w_s(\omega)=\frac{\kappa_s\nu_s}{p}(v_s(\omega)-\sum_{r\in
s}e^{-\omega T_{r}}u_r(\omega)),$$
\begin{equation}\label{eq:RVLocalL3}
\begin{aligned}
\omega\bar{v}_s(\omega)=\frac{\rho_s}{p}\left({\gamma\sum_{r\in
s}x_r^{-\frac{1}{p}}e^{-\omega
T_r}u_r(\omega)}\right.\\
\left.{+(1-\gamma)y_s^{-\frac{1}{p}}v_s(\omega)-\bar{y}_s^{-\frac{1}{p}}\bar{v}_s(\omega)}\right),
\end{aligned}
\end{equation}
$$u_r(\omega)=-px_r\left(a_s\bar{v}_s(\omega)+\sum_{j\in
r}e^{-\omega
T_{jr}}\frac{w_j(\omega)}{\lambda_r-\nu_s}-\frac{w_s(\omega)}{\lambda_r-\nu_s}\right),$$
$$ v_s(\omega)=-py_s\left(a_s\bar{v}_s(\omega)+\frac{w_s(\omega)}{\nu_s}\right).$$
By \eqref{eq:RVLocalL3}, we have
$$\begin{array}{ll}
 &(\omega+\sigma_s)\bar{v}_s(\omega)\\
=&\frac{\rho_s}{p}\left(\sum_{r\in s}\gamma
x_r^{-\frac{1}{p}}e^{-\omega
T_r}u_r(\omega)+(1-\gamma)y_s^{-\frac{1}{p}}v_s(\omega)\right),
\end{array}$$
where $\sigma_s=\rho_s\bar{y}_s^{-\frac{1}{p}}/p$.

We calculate that
$$
\left(\begin{array}{l} \bar{\bv}(\omega)\\  \bw(\omega)\end{array}\right)=-\bG(\omega) \left(\begin{array}{l} \bar{\bv}(\omega)\\  \bw(\omega)\end{array}\right).
$$
The matrix $\bG(\omega)$ is called the {\em return ratio} for $(\bar{\bv}, \bw)$ and
$$\begin{array}{l}
G_{s's'}(\omega)=\frac{\rho_{s'}a_{s'}}{(\omega+\sigma_{s'})}(\sum_{r\in s'}\gamma x_r^{\frac{1}{q}}e^{-\omega T_r}+(1-\gamma)y_{s'}^{\frac{1}{q}}),\\
G_{s'(S+s)}(\omega)=\frac{\rho_{s'}}{\omega+\sigma_{s'}}(-\sum_{r\in s} \frac{\gamma x_r^{\frac{1}{q}}}{\lambda_r-\nu_s}e^{-\omega T_r}\\
\hspace{4cm} +(1-\gamma)\frac{y_s^{\frac{1}{q}}}{\nu_s}), s=s'\\
G_{s'(2S+j)}(\omega)=\frac{\rho_{s'}}{\omega+\sigma_{s'}}\sum_{r\in s', j\in r}\frac{\gamma x_r^{\frac{1}{q}}}{\lambda_r-\nu_{s'}}e^{-\omega (T_r+T_{jr})}, \\
G_{(S+s)s'}(\omega)=\frac{\kappa_sa_s\nu_s}{\omega }(y_s-\sum_{r\in s}x_re^{-\omega T_r}), s=s'\\
G_{(S+s)(S+s)}(\omega)=\frac{\kappa_s\nu_s}{\omega}(\frac{y_s}{\nu_s}+\sum_{r\in s}\frac{x_r}{\lambda_r-\nu_s}e^{-\omega T_r}), \\
G_{(S+s)(2S+j)}(\omega)=-\frac{\kappa_s\nu_s}{\omega}\sum_{r\in s, j\in r}\frac{x_r}{\lambda_r-\nu_s}e^{-\omega (T_r+T_{jr})}, \\
G_{js'}(\omega)=\frac{\kappa_j\mu_ja_{s'}}{\omega}\sum_{j\in r, r\in s'}x_re^{-\omega T_{rj}}, \\
G_{j(S+s)}(\omega)=-\frac{\kappa_j\mu_j}{\omega}\sum_{j\in r, r\in s}\frac{x_r}{\lambda_r-\nu_s}e^{-\omega T_{rj}}, \\
G_{j(2S+j')}(\omega)=\frac{\kappa_j\mu_j}{\omega}\sum_{j\in r,
j'\in r}\frac{x_r}{\lambda_r-\nu_s}e^{-\omega (T_{rj}+T_{j'r})},
\end{array}$$
and all other entries are 0.

Let $\bar{\bG}$ be an $(2S+J)\times (2S+J)$ matrix with
$$\begin{array}{l}
\bar{G}_{s's'}(\omega)=\frac{\rho_{s'}a_{s'}}{(\omega+\sigma_{s'})}(1-\gamma)y_{s'}^{\frac{1}{q}},\\
\bar{G}_{s'(S+s)}(\omega)=\frac{\rho_{s'}}{\omega+\sigma_{s'}}(1-\gamma)\frac{y_s^{\frac{1}{q}}}{\nu_s}, s=s'\\
\bar{G}_{(S+s)s'}(\omega)=\frac{\kappa_s\nu_sa_s}{\omega}y_s, s=s'\\
\bar{G}_{(S+s)(S+s)}(\omega)=\frac{\kappa_s}{\omega}y_s
\end{array}$$
and all other entries are 0. It can be verified that
$$\bG(\omega)=\bP\bY(\omega)\bR(-\omega)^{\rm T}\bX(\omega)\bR(\omega)\bP^{-1}+\bar{\bG}(\omega),$$
where $\bX(\omega)$ is an $R\times R$ diagonal matrix with entries
$X_{rr}(\omega)=e^{-\omega T_r}/(\omega T_r)$, $\bY(\omega)$ is an
$(2S+J)\times (2S+J)$ diagonal matrix with entries
$Y_{s's'}(\omega)=\frac{\omega}{\omega
+\sigma_{s'}},Y_{ss}(\omega)=1,Y_{jj}(\omega)=1$, and $\bP$ is an
$(2S+J)\times (2S+J)$ diagonal matrix with entries
$P_{s's'}=\left(\frac{\rho_{s'}}{a_{s'}\bar{y}_s^{\frac{1}{p}}U'_s}\right)^{\frac{1}{2}},
P_{ss}=(\kappa_s\nu_s)^{\frac{1}{2}},
P_{jj}=\left(\kappa_j\mu_j\right)^{\frac{1}{2}}$, and $\bR(\omega)$
is an $R\times (2S+J)$ matrix where
$$
\begin{array}{l}
R_{rs'}(\omega)=\left(\gamma x_r^{\frac{1}{q}}T_r\right)^{\frac{1}{2}}\left(\rho_{s'}a_{s'}\right)^{\frac{1}{2}}, r\in s'\\
R_{rs}(\omega)=-\left(\frac{x_rT_r}{\lambda_r-\nu_s}\right)^{\frac{1}{2}}\left(\kappa_s\nu_s\right)^{\frac{1}{2}}, r\in s\\
R_{rj}(\omega)=\left(\frac{x_rT_r}{\lambda_r-\nu_s}\right)^{\frac{1}{2}}\left(\kappa_j\mu_j\right)^{\frac{1}{2}}e^{-\omega
T_{jr}}, j\in r
\end{array}
$$
and all other entries are 0.
Since the open loop system
\eqref{eq:RVoiceD_1}-\eqref{eq:RVoiceD_3} is stable, we just need to
show that the eigenvalues of the return ratio $G(\omega)$, for
$\omega=i\theta$, do not encircle the point $-1$ from the
generalized Nyquist stability criterion. Now, these eigenvalues are
identical to those of
\begin{equation}\label{eq:RVLocalL9}
\bY(\omega)\bR(-\omega)^{\rm
T}\bX(\omega)\bR(\omega)+\bP^{-1}\bar{\bG}(\omega)\bP.
\end{equation}

If $\lambda$ is an eigenvalue of the $\bG(i\theta)$, then we can
find a unit vector $\bz$ such that
$$\lambda\bz=\bY(i\theta)\bR(i\theta)^*\bX(i\theta)\bR(i\theta)\bz+\bP^{-1}\bar{\bG}(i\theta)\bP\bz,$$
where $*$ represents the matrix conjugate. Thus
$$\begin{array}{l}
\lambda\bz^*\bY(i\theta)^{-1}\bz=\bz^*\bR(i\theta)^*\bX(i\theta)\bR(i\theta)\bz\\
                            \hspace{2.5cm}     +\bz^*\bY(i\theta)^{-1}\bP^{-1}\bar{\bG}(i\theta)\bP\bz.
\end{array}$$
If $\lambda$ is real, since the real parts of
$\bz^*\bY(i\theta)^{-1}\bz$ and
$\bz^*\bY(i\theta)^{-1}\bP^{-1}\bar{\bG}(i\theta)\bP\bz$ are 1 and 0
respectively, we have
$$\lambda={\rm Re}(\bz^*\bR(i\theta)^*\bX(i\theta)\bR(i\theta)\bz).$$
Let $\bd=\bR(i\theta)\bz$. Then, since $\bX$ is diagonal,
$$\lambda=\sum_{r}|d_r|^2{\rm Re}(X_{rr}(i\theta))=\sum_{r}|d_r|^2{\rm Re}\left(\frac{e^{-i\theta T_r}}{i\theta T_r}\right).$$
Since ${\rm Re}\left(\frac{e^{-i\theta T_r}}{i\theta T_r}\right)\ge
-\frac{2}{\pi}$ for all $\theta$ \cite{Vinnicombe02}, and hence
$\lambda\ge (-2/\pi)K$, where $K=\|\bR(i\theta)\bz\|^2$.

Next, we bound $K$. Let $Q$ be the $(2S+J)\times (2S+J)$ diagonal
matrix taking values
$Q_{s's'}=\sqrt{\frac{\bar{y}_s^{\frac{1}{p}}U'_s}{\rho_sa_s}}$,
$Q_{ss}=\sqrt{\frac{\nu_s}{\kappa_s}}$ and
$Q_{jj}=\sqrt{\frac{\kappa_j}{\mu_j}}$. Let $\rho(\cdot)$ denote the
spectral radius, and $\|\cdot\|_\infty$ the maximum row sum matrix
norm. Then
$$\begin{array}{ll}
K &=\bz^*\bR(i\theta)^*\bR(i\theta)\bz\\
  &\le\rho(\bR(i\theta)^*\bR(i\theta))\\
  &=\rho(\bQ^{-1}\bR(i\theta)^*\bR(i\theta)\bQ)\\
  &\le \|\bQ^{-1}\bR(i\theta)^*\bR(i\theta)\bQ\|_{\infty}\\
  &<\frac{\pi}{2},
  \end{array}  $$
the last inequality follows from
\eqref{eq:RVLocalS22} and \eqref{eq:MVlocalstability1}-\eqref{eq:MVlocalstability3}.

So we have that $\lambda>-1$ for any real eigenvalue $\lambda$. Thus, when the loci of the eigenvalues of $\bG(i\theta)$ for $-\infty<\theta<\infty$ cross the real axis, they do so to the right of -1. Hence the loci of the eigenvalues of $\bG(i\theta)$ cannot encircle -1, the generalized Nyquist stability criterion is satisfied and the system \eqref{eq:RVoiceD_1}-\eqref{eq:RVoiceD_5} is stable, in the sense that $v_s(t)\to 0, \bar{v}_s(t)\to 0, w_j(t)\to 0$ exponentially, for all $s,j$, as $t\to \infty$. There remains the possible that $\Mu(t)$ might hit a boundary of the positive orthant, and invalidate the linearization \eqref{eq:RVLocalS1}-\eqref{eq:RVLocalS3}. To rule out this possibility, note that there exists an open neighborhood of $\Mu$, say $\mcN$, such that $\Mu(t)>0, t\in (-T_{\max},0)$, the linearization is valid. Thus $\mcN$ is as required.

Finally we shall relax the assumption that $\mu_j>0$ for all $j$. Since $(\Mu, \Nu, \bar{\by})$ be an equilibrium point of system
\eqref{eq:RVoiceD_1}-\eqref{eq:RVoiceD_5} with no almost saturated links, $\mu_j=0$ implies $\dot{\mu}_j(t)<0$. Thus there is a neighborhood of $\Mu$, say $\mcM$, such that, on $\mcM$, the linearization of \eqref{eq:RVoiceD_1}-\eqref{eq:RVoiceD_5} coincides with the case where we discard all $j$ such that $\mu_j=0$. Therefore, as above, we may choose an open neighborhood $\mcN\subset \mcM$ such that for any initial trajectory $((\Mu(t), \Nu(t),
\bar{\by}(t)), t\in (-T_{\max},0))$ with $\Mu(t)$ lying within the neighborhood
$\mcN$, $(\Mu(t),\Nu(t))$ converge as $t\to \infty$ to the solution
$(\Mu, \Nu)$ to the optimization problem \eqref{eq:NewDual} and
$(\bx(t),\by(t),\bar{\by}(t)^{\frac{1}{q}})$ converge as $t\to
\infty$ to the solution $(\bx, \by, \bar{\by}^{\frac{1}{q}})$ to the
optimization problem \eqref{eq:NewPrimal}.
\done

%

\end{document}